\documentstyle[11pt,dina4p]{article}
\def\HH{{\cal H}}
\def\RR{{\cal R}}
\def\RC{{\cal R}^c}
\def\OM{O_M}
\def\OMP{O_{M'}}
\def\OMC{O_{M^c}}
\def\PA{\psi_A}
\def\PG{\psi_G}
\def\PS{\psi_S}
\def\PT{\psi_T}
\def\PX{\psi_X}
\def\PY{\psi_Y}
\def\bfx{{\bf x}}
\def\bfy{{\bf y}}
\def\be{\begin{equation}}
\def\ee{\end{equation}}
\title{There are No Causality Problems for Fermi's Two Atom System} %
\author{Detlev Buchholz$^{a}$ \hspace{0.1cm} and \hspace{0.1cm} Jakob
Yngvason$^{b}$ \\[0.5cm]
${}^a$ II. Institut f\"ur Theoretische Physik, Universit\"at Hamburg \\
Luruper Chaussee 149, D-22761 Hamburg, Germany \\[0.2cm] %
${}^b$ Science Institute, University of Iceland \\ Dunhaga 3, IS-107
Reykjavik, Iceland}
\date{DESY 94-027 \\ hep-th/9403027}
\begin{document}
\maketitle
\begin{abstract}
A repeatedly discussed gedanken experiment, proposed by Fermi to check
Einstein causality, is reconsidered. It is shown that, contrary to a
recent statement made by Hegerfeldt, there appears no causality paradoxon
in a proper theoretical description of the experiment.
\smallskip

\end{abstract}

\noindent
In a recent letter \cite{H} Hegerfeldt discusses a gedanken experiment
proposed by Fermi to determine the speed by which causal influences
propagate. He argues that the theoretical description of this experiment
in terms of transition probabilities leads to results which are in
conflict with the existence of a maximal propagation speed $c$. Hegerfeldt
suggests that the difficulties might disappear if one drops some implicit
assumptions about the preparability of states with certain specific
localization properties (points a) to c) of his conclusions).
He does not settle the question whether the theory complies with Einstein
causality, however.

In this letter we would like to set forth that there are no difficulties
with Einstein causality in the theoretical setting of relativistic quantum
field theory (RQFT). We will explain why, on one hand, transition
probabilities are not a suitable tool for a thorough discussion of causal
effects: what is required is a comparison of expectation
values. On the other hand we will show that the points indicated by
Hegerfeldt as possible loopholes to evade causality problems have to be
taken seriously indeed and require a more careful analysis. Taking these
facts into account we will arrive at the conclusion that there is no
conflict between the gedanken experiment of Fermi and the theoretical
predictions of RQFT.

The experimental setup envisaged by Fermi to determine the propagation
speed $c$ of causal influences can be described as follows (cf. \cite{F}
and, for further references, \cite{H}): one should prepare a state
consisting of two atoms which are localized in disjoint regions separated
by a distance $R$. One atom should be in its ground state, the other one
in an excited state. If causal influences propagate with maximal velocity
$c$ one should not observe any impact of the excited atom on the atom in
the ground state (e.g., by an emitted
photon) within the time interval $0 < t < R/c$. Such events should be
observed only at later times.

It is of importance that the atoms in this experiment have well defined
localization properties. Hence in a theoretical discussion of the setup
one first has to define in precise terms what one means by the statement
that some physical state $S$ (e.g., the state considered by Fermi,
consisting of two atoms ) looks at time $t=0$, say, inside a region $\RR$
like a given state $G$ (e.g., like an atom in its ground state) \cite{R1}.
 From the point of view of physics the appropriate definition seems
to be the following one: it is impossible to distinguish $S$ from $G$ by
any measurement $M$ which one performs at the given time in the region
$\RR$. In the theoretical setting this amounts to the requirement that the
expectation values of all operators (observables) $O_M$ corresponding to
these measurements have to coincide. Hence if $\PS$, $\PG$ denote the
Hilbert space vectors representing $S$ and $G$ respectively there must hold
\be \langle \PS | \OM | \PS \rangle \, = \, \langle \PG | \OM | \PG
\rangle. \ee
This condition on $\PS$ involves matrix elements of a multitude of
observables and one may therefore ask whether it can be reformulated in
terms of projection operators which test whether $S$ coincides with $G$ in
the given region. The answer turns out to be different in relativistic and
non-relativistic theories where the following alternatives hold.

i) Linear combinations of vectors $\PS$ satisfying (1) again satisfy this
condition (after normalization). This case is generic in non-relativistic
quantum field theory, where it appears for a total set of states $G$.
These states describe a situation where one has locally maximal
information about the underlying system (locally pure states). Examples
are the Fock vacuum and all coherent states. For any such state $G$ one
may consider the projection operator $O_{G \, \mbox{{\scriptsize inside}}
\, \RR \mbox{{\scriptsize ?}}}$, projecting onto the subspace of the
physical Hilbert space spanned by all vectors $\PS$ satisfying condition
(1). This projection operator can be used to decide whether some arbitrary
state $A$ coincides with $G$ inside of $\RR$: one simply has to calculate
the transition probability $\langle \PA | O_{G \, \mbox{{\scriptsize
inside}}\, \RR \mbox{{\scriptsize ?}}}| \PA \rangle $ and to check
whether it is equal to $1$. Thus in the non-relativistic setting there
exist projection operators which completely fix the local properties of
states and it is then possible to study these properties in terms of
transition probabilities.

ii) Certain normalized linear combinations of vectors $\PS$ satisfying
condition (1) do not comply with this condition. This is the case in RQFT
for {\em every} choice of $G$. Phrased differently: all states $G$ look
locally like mixtures (of an, as a matter of fact, infinite number of
states). In contrast to the non-relativistic case, it is thus not possible
to fix the local properties of states with the help of projection
operators \cite{R2}.

In view of these facts one is forced to base the local analysis of states,
which is fundamental in any discussion of causal effects, on a comparison
of states in the sense of relation (1); transition probabilities are not
the adequate tool to study this issue in relativistic theories. This
important point may be illustrated by a simple example. If $G$ is, e.g.,
the vacuum state in RQFT, then the expectation value $ \langle \PG | \OM |
\PG \rangle $ cannot vanish for any positive operator $O_M$
corresponding to a localized measurement \cite{R3}. To test for a local
deviation of $S$ from $G$ one can therefore not take as a criterion that
the expectation value of some suitable projection operator (or, more
generally, some positive operator) has a non-zero expectation value in the
state $S$. For this expectation value would be non-zero even if the vacuum
$G$ is present. This point has been overlooked in \cite{H} and led to
the apparent paradoxa. A deviation of $S$ from $G$ would show up,
however, in {\em different} values of the
left and right sides of (1) for some $O_M$ \cite{R4}.

After these general remarks let us turn now to the actual discussion of
Fermi's gedanken experiment. Let $X$ be the ground state of an isolated
atom which is localized in the vicinity of $0$ and surrounded by vacuum
and let $\PX$ be the corresponding state vector. Following Fermi, we
consider a
state $S$, described by a vector $\PS$, which looks inside a ball $\RR$ of
radius $R$ about $0$ like $X$, i.e., \be \langle \PS | \OM | \PS \rangle
\, = \, \langle \PX | \OM | \PX \rangle \ee
for all observables $\OM$ which are localized in $\RR$. In the complement
$\RC$ of this ball $S$ may look like any other state $Y$, e.g., like some
excited atom.
If, as expected, the subsystem in $\RC$ does not affect the atom in $\RR$
within the time interval $0 < t < R/c$ it should not be possible to
discriminate $S$ from $X$ by any measurement $M'$ which one performs at
time $t$ within the ball $\RR_t$ of radius $R - c t$ about $0$. Phrased
differently: $S$ should still look like an atom in its ground state within
the smaller region $\RR_t$. Hence, using the
Heisenberg picture, there should hold in the theoretical setting \be
\langle \PS | \OMP (t) | \PS \rangle \, = \, \langle \PX | \OMP (t) | \PX
\rangle, \ee where
\be \OMP (t) \, = \, e^{i t H / \hbar} \OMP e^{- i t H / \hbar}. \ee It is
a fundamental fact that relation (3) is a consequence of relation (2) in
theories where the underlying equations of motion are hyperbolic.
Within the setting of RQFT this fact is called ``primitive causality''
\cite{HS} and has been established in models, cf. for example \cite{GJ}.
It is independent of the spectral properties of the generator
$H$, which in fact depend on the systems which one considers (few body
systems, thermal states, etc.). Hence in this respect the predictions of
RQFT are in perfect agreement with the ideas of Fermi.

There remains, however, the question of whether the theory is capable of
describing the physical situation envisaged by Fermi, cf. point c) in
\cite{H}. Given two vectors $\PX$, $\PY$ corresponding to states $X$, $Y$,
does there exist a vector $\PS$ describing the {\em composite} state $S$
which looks like $X$ in a given region $\RR$ and like $Y$ in its complement
$\RC$? These requirements fix $\PS$ completely and can be cast into the
following condition on the expectation values, \be \langle \PS | \OM \cdot
\OMC | \PS \rangle = \langle \PX | \OM | \PX \rangle
\langle \PY | \OMC | \PY \rangle, \ee
where $\OM$ and $\OMC$ respectively denote operators corresponding to
measurements in $\RR$ and $\RC$. Relation (5) gives formal expression to
the idea that $S$ is composed of states $X$ and $Y$ which are localized
(in the sense of condition (1)) in disjoint regions and do not
``overlap'', cf. point a) in \cite{H}.

The question of whether such product states exist is known in RQFT as the
problem of ``causal (statistical) independence'' \cite{HK,Sch}. It has an
affirmative answer \cite{R}, but the vectors $\PS$ have in general
infinite energy even if $\PX$ and $\PY$ have finite energy. This
phenomenon can be traced back to the uncertainty principle and may be
easily understood in the framework of non-relativistic quantum mechanics:
if $\PX$ and $\PY$ are the configuration space wave functions of
distinguishable systems then the wave function $\PS$ of the composite
state is given by
\be \PS ( \bfx , \bfy ) = N \cdot
\left\{ \begin{array}{c@{{} \quad }l} \PX ( \bfx ) \PY ( \bfy ) &
\mbox{for $\bfx \in\RR$, $\bfy\in\RR^c$} \\ 0 & \mbox{otherwise}
\end{array} \right. \ee
where $N$ is a normalization constant. This function has in general a
discontinuity when $\bfx$ or $\bfy$ are at the boundary of $\RR$, unless
the wave functions $\PX$ and $\PY$ happen to vanish at these points. (Note
that wave functions of states with sharp energy, such as bound states, in
general do not have such nodes.) As a consequence, the expectation value
of the Hamiltonian becomes infinite. In RQFT the situation is even worse
because of pair creation. There it turns out that due to such processes
the vector $\PS$ cannot be an element of the physical Hilbert space
$\HH$ describing few body systems (e.g., Fock space in free field theory).

Thus the theory predicts that every member of the ensemble described by
$S$ has infinite energy. Hence a preparation of this state would not be
possible in practice. This fact seems to be in conflict with the ideas of
Fermi, but the apparent difficulty disappears if one notices that for the
determination of $c$ it suffices to consider ``tame'' states $T$ which
look like $X$ in a region ${\RR}_{<}$ and like $Y$ in ${\RR}_{>}^{c}$,
where $\RR_{<}$ is slightly smaller and ${\RR}_{>}$ slightly larger than
$\RR$. It is not really necessary to completely fix the state $T$ in the
layer between these two regions. By making this layer sufficently small
one can then determine $c$, as outlined above, with arbitrary precision.

It has been shown in RQFT under very general conditions that there exist
vectors $\PT$ in the physical Hilbert space $\HH$ which satisfy condition
(5) for the slightly smaller regions \cite{BW,BDF}. The existence of such
vectors has also been established in models \cite{B,D,S}. Thus also
in this respect Fermi's gedanken experiment poses no theoretical problems.

It should be mentioned that the vectors $\PT$ have to be carefully
adjusted in the layer between the two regions ${\RR}_{<}$ and
${\RR}_{>}^{c}$ in order to become elements of the physical Hilbert space
$\HH$. This fact may be viewed as the process of ``renormalization'',
indicated in point b) of \cite{H}, which surrounds
state $X$ by some ``cloud''. In more physical terms: any state of the type
considered by Fermi which can actually be prepared in an experiment
necessarily contains, besides the two atoms, other particles, e.g.,
photons. This fact is the basic reason for the apparently non-causal
behaviour of transition probabilities, discussed in \cite{H}. But it is
not in conflict with the existence of a maximal propagation speed $c$.


\begin{thebibliography}{99}
\bibitem{R1} We
assume that $G$
exists as a global state described by a vector in the physical Hilbert
space. This assumption causes no problems if one considers configurations
of atoms. But
it would require further discussions if one thinks, e.g., of quarks as
subsystems of hadrons (confinement).
\end{thebibliography}
\enddocument